\documentclass{PoS}

\usepackage{amsmath, amssymb, multirow}
\usepackage{setspace}

\newcommand{\hess}{H.E.S.S.}
\newcommand{\gr}{$\gamma$-ray}
\newcommand{\grs}{$\gamma$-rays}

\title{The H.E.S.S. multi-messenger program}

\ShortTitle{The H.E.S.S. multi-messenger program}

\author{\speaker{Fabian Sch\"ussler}\\
        CEA-Saclay / Irfu, Saclay, France \\ 
        E-mail: \email{fabian.schussler@cea.fr}}
\author{ Arnim Balzer, Francois Brun, Pierre Brun, Wilfried Domainko, Matthias Fuessling, Clemens Hoischen, Gerd Pühlhofer, Anita Reimer, Gavin Rowell on behalf of the H.E.S.S. Collaboration}        

\abstract{Based on fundamental particle physics processes like the production and subsequent decay of pions in interactions of high-energy particles, close connections exist between the acceleration sites of high-energy cosmic rays and the emission of high-energy gamma rays and high-energy neutrinos. In most cases these connections provide both spatial and temporal correlations of the different emitted particles. The combination of the complementary information provided by these messengers allows to lift ambiguities in the interpretation of the data and enables novel and highly sensitive analyses. \\
In this contribution the H.E.S.S. multi-messenger program is introduced and described. The current core of this newly installed program is the combination of high-energy neutrinos and high-energy gamma rays. The search for gamma-ray emission following gravitational wave triggers is also discussed. Furthermore, the existing program for following triggers in the electromagnetic regime was extended by the search for gamma-ray emission from Fast Radio Bursts (FRBs). An overview over current and planned analyses is given and recent results are presented.\\[4ex]
}

\FullConference{The 34th International Cosmic Ray Conference,\\
		30 July- 6 August, 2015\\
		The Hague, The Netherlands}

\begin{document}
\section{Introduction}
The key question to resolve the long standing mystery of the origin of high-energy cosmic rays is to locate the sources and study the acceleration mechanisms able to produce fundamental particles with energies orders of magnitude above man-made accelerators. The combination of complementary information provided by multiple messengers and novel techniques will increase the chances to achieve this century old task. \\
Thanks to the increase in computing power combined with advances in electronics and data analysis techniques, real-time analyses of data taken by astrophysical observatories at all wavelengths are now becoming possible. In addition, new messengers provide additional information on high-energy astrophysical processes. Due to fundamental particle physics processes like the production and subsequent decay of pions in interactions of high-energy particles, the acceleration sites of high-energy cosmic rays might also be sites of gamma ray and neutrino production. Unlike cosmic rays, both gamma rays and neutrinos are neutral messengers and therefore point back to their source. If produced within the accessible horizon (which is limited for gamma rays due to pair creation processes on the extragalactic background light) spatial and temporal correlations of the different emitted particles could exist. The \hess\ multi-messenger program will fully exploit these correlations and might allow opening a new window to the high-energy universe: real-time multi-messenger astronomy at TeV energies.

\section{High-energy neutrinos}
The core of the \hess\ multi-messenger program is to exploit the intimate connection between high-energy neutrinos and \grs. Provided appropriate conditions of the environment of cosmic accelerators (e.g. magnetic fields, matter and field densities, etc.), high-energy (hadronic) particles are potentially undergoing interactions with matter and radiations fields within and/or surrounding the acceleration sites. The light mesons, predominately pions, created in these interactions will decay by emitting both high-energy neutrino as well as \grs.

\begin{figure*}[!h]
\centering
\includegraphics[width=0.63\textwidth]{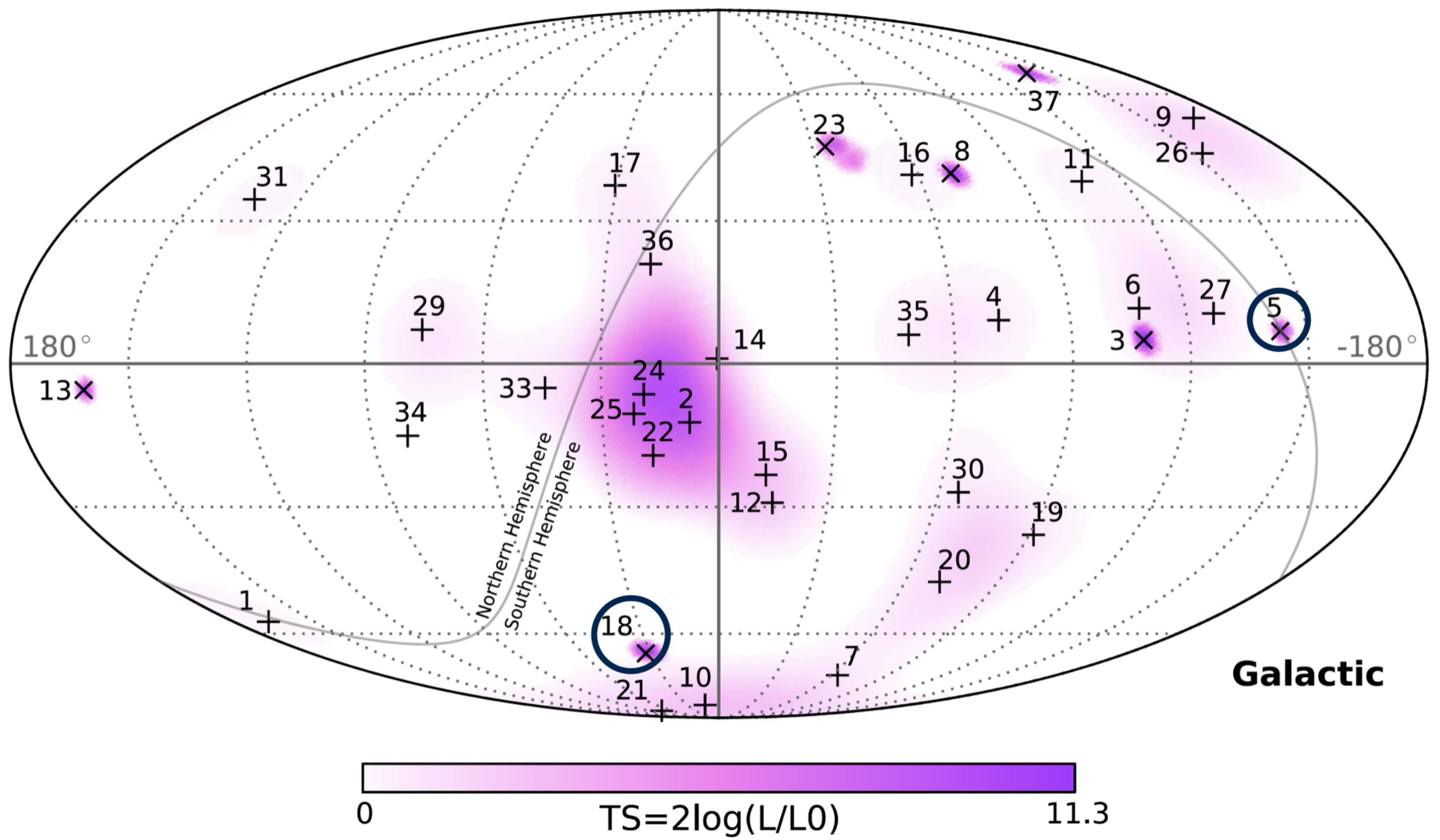}
\caption{Arrival directions in Galactic coordinates of high-energy neutrino events detected by IceCube. Here we study the regions around events IC-5 and IC-18 (highlighted by the black circles) . Modified from~\cite{IC_HESE}.}\label{fig:ICsky}
\end{figure*}

Two major instruments searching for these astrophysical neutrinos are currently in operation: IceCube at the South Pole~\cite{IceCube}, complemented by ANTARES~\cite{ANTARES}, a smaller detector in the Mediterranean Sea. Despite tremendous efforts, so far none of the neutrino telescopes has found any significant localized excess (e.g.~\cite{IC_pointsources}). Yet, a significant breakthrough in the search for astrophysical neutrinos has been made very recently by the IceCube collaboration. In three years of data, IceCube was able to single out 37 neutrinos with energies in the range of 30 TeV to 2 PeV that interacted within the instrumented volume (see Fig.~\ref{fig:ICsky}). The atmospheric background contribution has been estimated as $8.4 \pm 4.2$ from cosmic ray muon events and $6.6^{+5.9}_{-1.6}$ atmospheric neutrinos events, which yields a statistical significance of $5.7 \sigma$~\cite{IC_HESE}. 
However, the origin of these neutrinos is unknown and no significant clustering or excess at small angular scales has been found so far. 


Here we present detailed studies of the regions around some of the neutrino events detected by IceCube, focussing on track-like events that provide a good angular resolution of the order of $1^\circ$. The studied regions are highlighted in Fig.~\ref{fig:ICsky}. The main aim of the presented analysis is the search for steady gamma-ray emitters in the vicinity of the arrival direction of the observed high-energy neutrino events, the discovery of which would trigger detailed follow-up studies at various wavelengths and messengers to clarify a potential association. The two selected regions around events with the identifiers IC-5 and IC-18 (see~\cite{IC_HESE}) have been studied in the GeV \gr\ domain using data from the the LAT instrument onboard the Fermi satellite and data taken by the \hess\ Imaging Air Cherenkov Telescopes (IACTs), which are sensitive to multi-GeV-TeV \grs.

\begin{figure*}[!t]
\vspace{-3mm}
\centering
\includegraphics[width=0.72\textwidth]{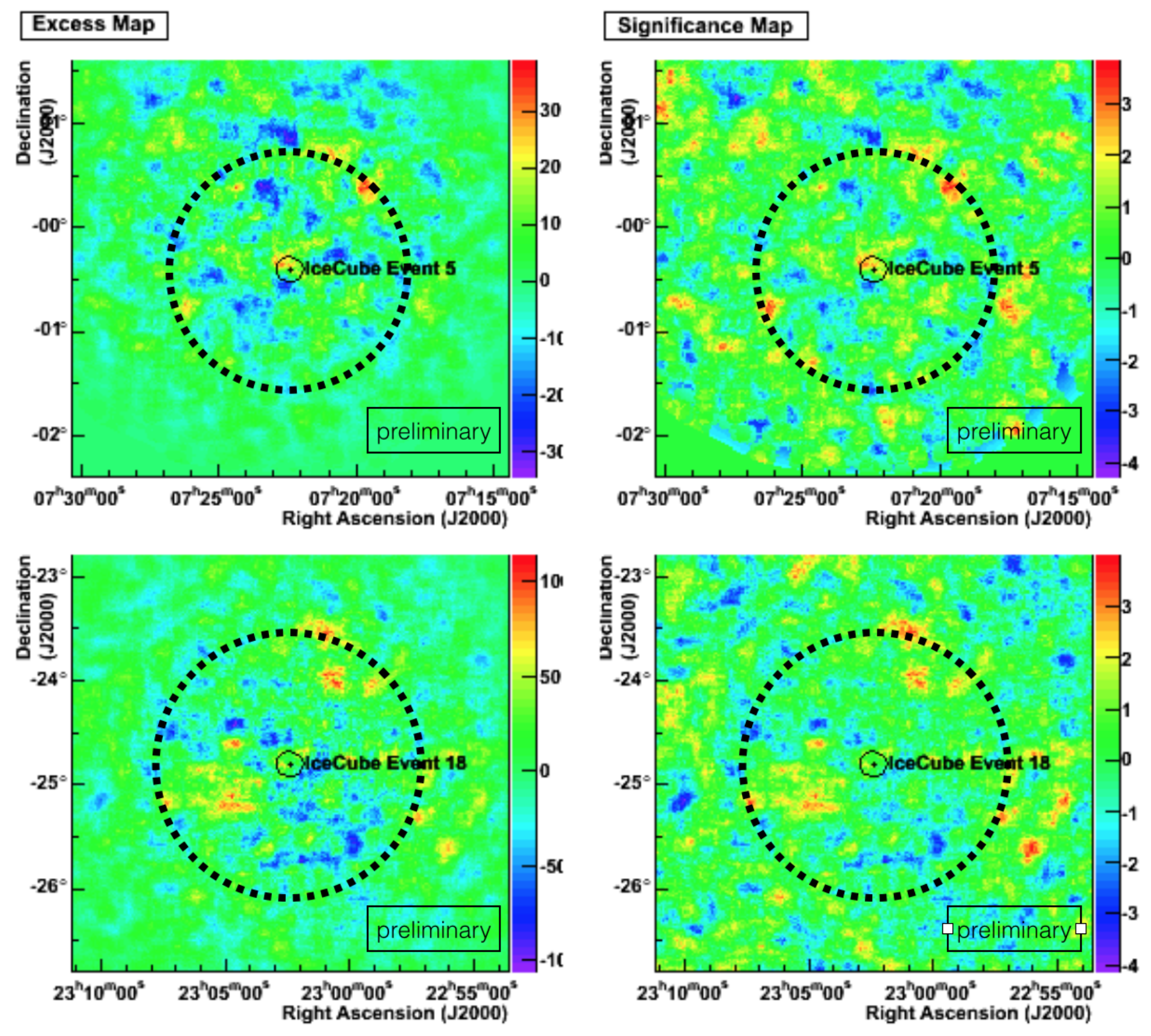}
\caption{\hess\ observations of IceCube events 5 and 18. The dashed circles denotes the individual median angular uncertainty on the neutrino directions provided by IceCube~\cite{IC_HESE}. For comparison, the inner, solid circles show the approximate size of a point-like source after convolution with the \hess\ angular resolution. Left plots: Map of VHE gamma-ray events exceeding the background expectation. Right plots: Point source significance map.}\label{fig:HESS}
\end{figure*}
  
\subsection{\hess\ analyses}
Dedicated observations on the two IceCube events IC-5 and IC-18 were taken with the \hess\ IACTs during 2014.\\ The region around IC-5 has been observed for about $2~\mathrm{h}$ at a zenith angle of $30^\circ$ in monoscopic mode with the H.E.S.S.-II 28m telescope. After correcting for acceptance effects the effective live time corresponds to $0.9~\mathrm{h}$. The data were analyzed using the Model Analysis~\cite{ParisAnalysis} with standard gamma-hadron separation and event selection cuts and using a background estimation exploiting the uniformity of the acceptance across the field-of-view (FoV) of the system. No significantly enhanced gamma-ray flux has been detected. The distribution of gamma-ray events exceeding the background is shown for the full ROI in the left plot of Fig.~\ref{fig:HESS}. The right plot of Fig.~\ref{fig:HESS} shows the map of the Li \& Ma significances~\cite{LiMa}. They are both fully compatible with the background expectation.\\
The region around the neutrino event IC-18 has been observed for almost $12~\mathrm{h}$ (effective live time after acceptance correction of $9.5~\mathrm{h}$) with the full array of the five \hess\ telescopes at zenith angles around $10^\circ$. The data were again analyzed using the Model Analysis~\cite{ParisAnalysis} with standard gamma-hadron separation and event selection cuts. Thanks to the high available statistics the background could be determined using the "reflected background" method described in~\cite{RingBg}, a method that exploits the properties of the wobble data taking mode and yields very low systematic uncertainties related to the acceptance of the camera system. No significantly enhanced gamma-ray flux has been detected and the resulting maps (cf. Fig.~\ref{fig:HESS}) are fully compatible with the background expectation.  \\

\begin{figure*}[!t]
\vspace{-3mm}
\centering
\includegraphics[width=0.92\textwidth]{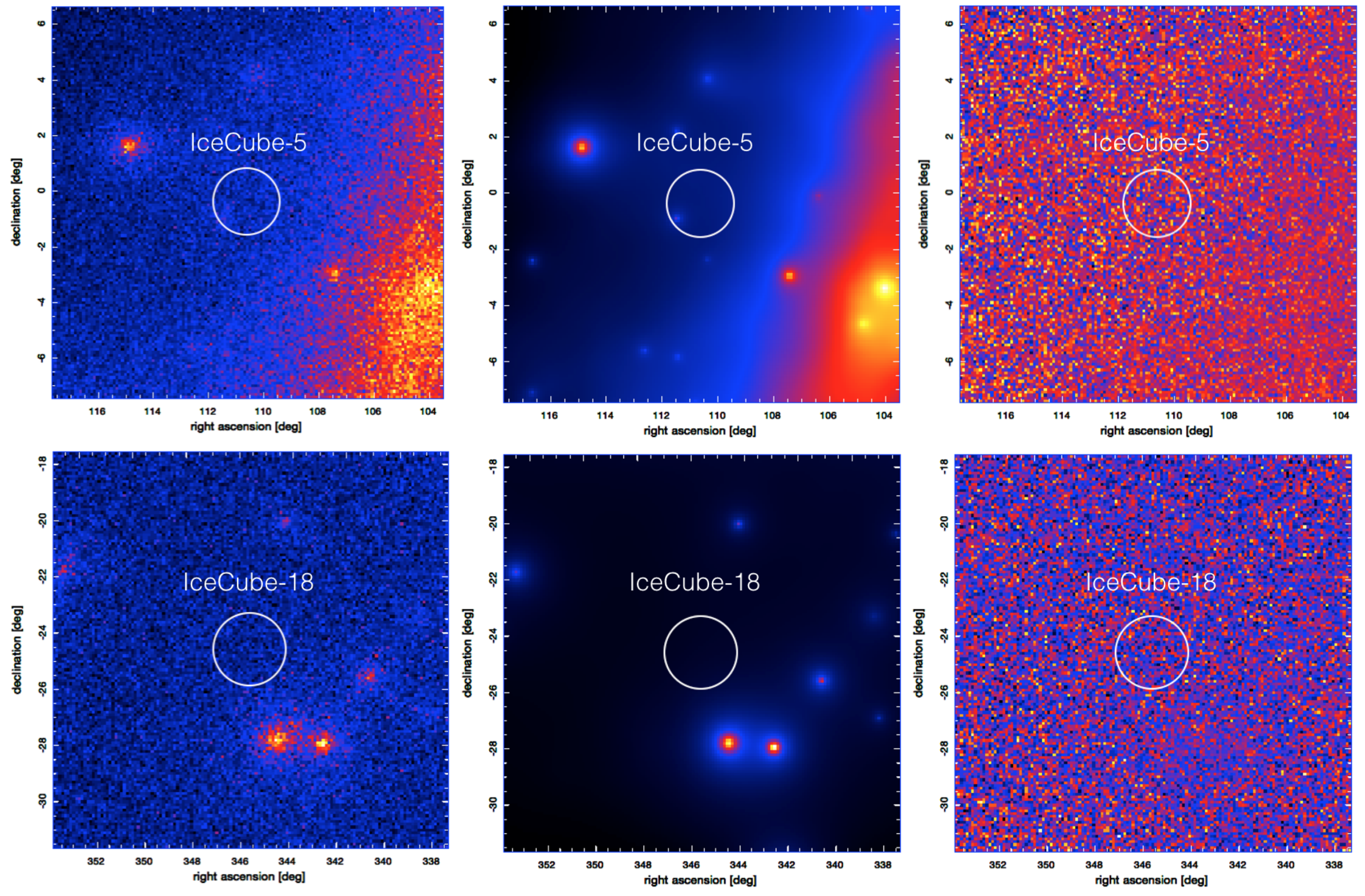}
\caption{Region around the studied IceCube events as seen by the Fermi-LAT gamma-ray observatory. Left plots: Count maps of \grs\ detected by Fermi-LAT in equatorial coordinates. Center plots: Maps showing the model of \gr\ sources fitted to the Fermi-LAT data. Right plots: Normalized residual maps. The white circles denotes the individual median angular uncertainty on the neutrino directions provided by IceCube~\cite{IC_HESE}.}\label{fig:Fermi}
\end{figure*}

\subsection{Fermi-LAT analyses}
Fermi-LAT data recorded between 2008-08-04 and 2015-05-01 has been analyzed (P7Rep) in the full available energy range (100 MeV - 300 GeV). All photon candidates (evclass=2) above $E=100~\mathrm{MeV}$ fulfilling the basic quality criteria proposed by the Fermi collaboration (quality=1; LAT config=1; rock angle$<52^\circ$; zenith$<100^\circ$) within a $10^\circ \times 10^\circ$ region of interest centered at the two locations of interest (IC-5: $\mathrm{RA}=110.6^\circ, \mathrm{Dec}= -0.4^\circ$ and IC-18: $\mathrm{RA}=345.6^\circ, \mathrm{Dec}= -24.8^\circ$) have been selected. The count maps of the selected gamma-ray events is shown in the left column of Fig.~\ref{fig:Fermi}. The gamma-ray emission observed in the Fermi-LAT data has been modeled using the information given in the 3FGL catalogue~\cite{3FGL}, fixing the parameters for known sources further away than $3~\mathrm{deg}$ from the center of the region of interest (ROI) to the 3FGL values. An additional point-like source with a power-law energy spectrum has been added to the description before fitting the model parameters to the count map of the selected events. The fit did not yield significant emission from the additional putative source in either of the studied fields (cf. Fig.~\ref{fig:Fermi}). It should be noted that the ROI of IceCube-5 (median angular resolution of $1.2^\circ$) comprises an AGN know to emit \grs\: PKS 0723-008 which is located $1.04^\circ$ from the center of the ROI. Given the high chance probability of finding a known \gr\ emitting AGN within the error-box of a track-like high-energy neutrino detected by IceCube of about $37~\%$ and based on flux level and spectrum of PKS 0723-008, an association with IceCube-5 can be discarded (e.g.~\cite{Antony2015}). 

\subsection{Upper limits on the \gr\ flux}
\begin{figure}[!t]
\vspace{-6mm}
  \centerline{
   \raisebox{-0.49\height}{\ \includegraphics[width=0.48\linewidth]{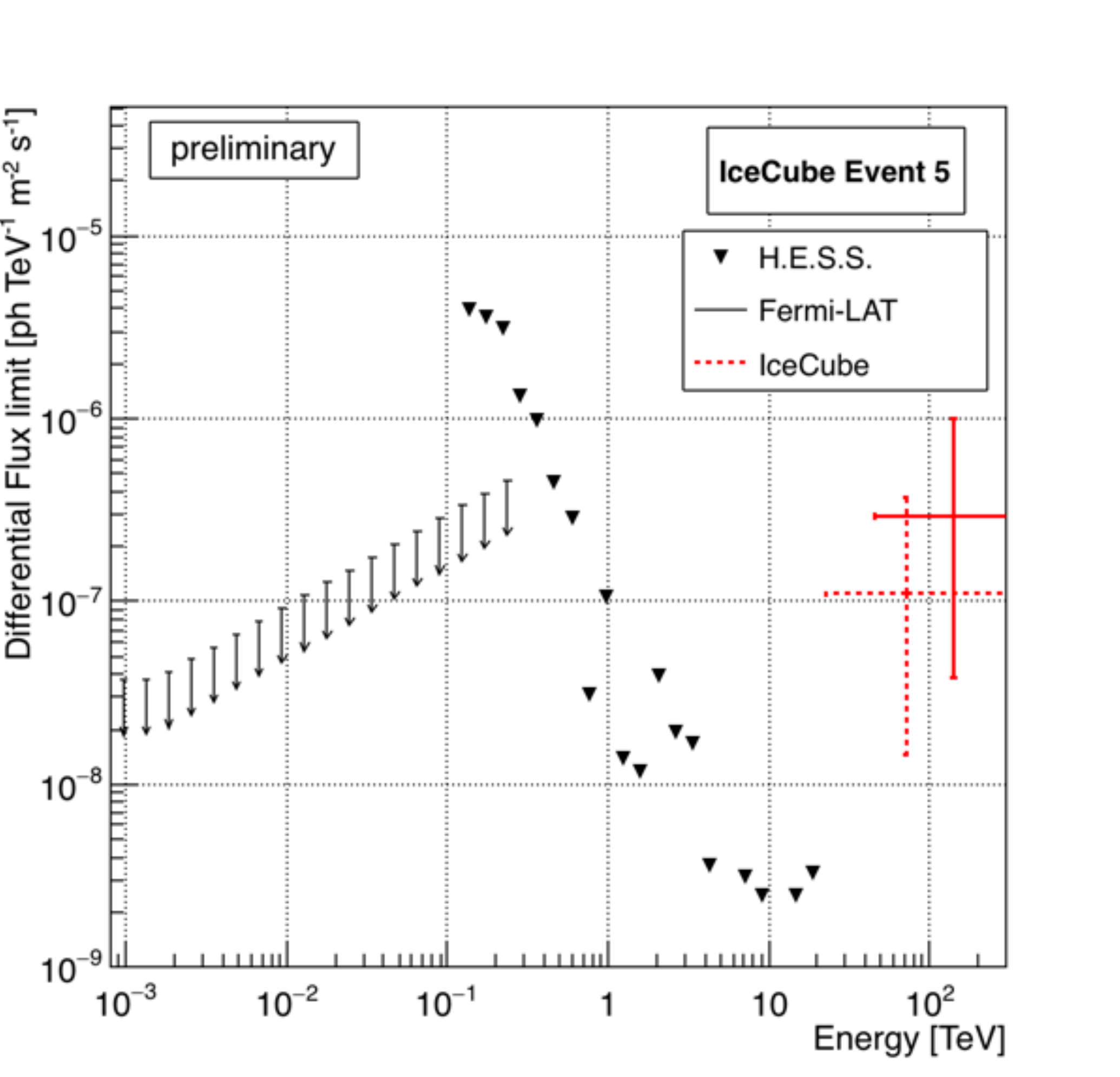}}
  \hfill
  \raisebox{-0.49\height}{\includegraphics[width=0.48\textwidth]{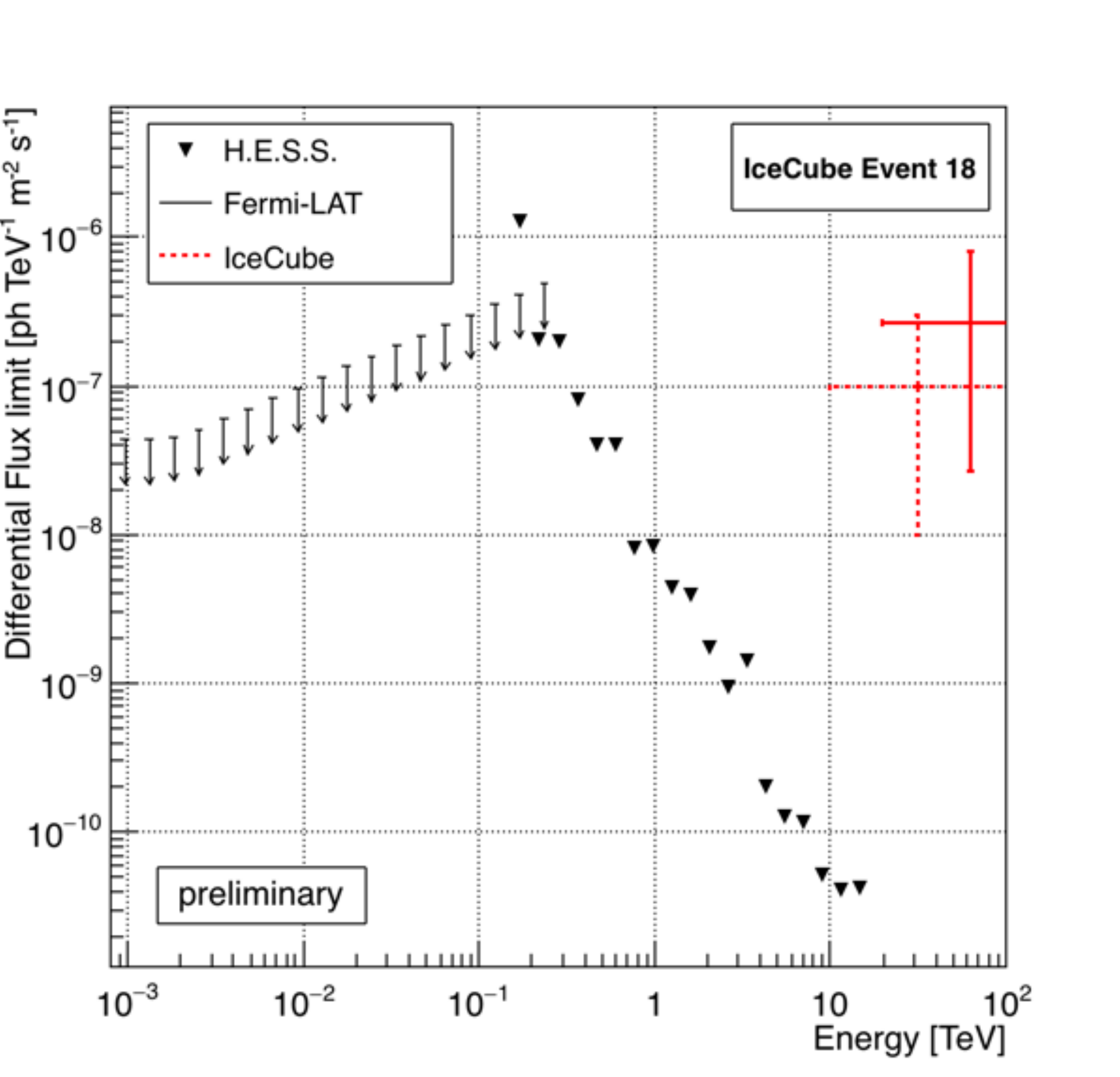}}
  }
   \caption{VHE gamma-ray flux limits $\Phi_\mathrm{UL}$ at 99~\% CL derived from the H.E.S.S. (black triangles) and Fermi-LAT (black arrows) observations assuming a point-like source with an $E^{-2}$ energy spectrum. The estimate of the \gr\ flux (solid, red marker) derived from the IceCube observation (dashed red marker) of event IC-5 (left plot) and IC-18 (right plot) is shown for reference. }
\label{fig:fluxlimits}	
 \end{figure}

Given the absence of a significant very high energy \gr\ signal in the observed regions upper limits on the VHE and HE \gr\ flux have been derived. Flux limits $\Phi_\mathrm{UL}$ have been calculated for a point-like source in the center of the ROI (see solid circles in Fig.~\ref{fig:HESS}) and for a generic $E^{-2}$ energy spectrum. Following the method introduced by Feldman \& Cousins~\cite{FeldmanCousins}, the obtained $99~\%$ confidence level limits from the H.E.S.S. observations are shown as black triangles in Fig.~\ref{fig:fluxlimits}. The limits on the high energy \gr\ flux derived from the Fermi-LAT data are shown as arrows. For reference, the red markers denote the expectation from the neutrino candidate events. The estimation of the neutrino flux related to the individual events (dashed markers) follows the approach described in~\cite{PadovaniResconi}, assuming that each single event comes from one astrophysical counterpart with the observed flux following an $E^{-2}$ energy spectrum and being spread over 1 dex in energy. The conversion into a \gr\ flux (solid red markers) relies on Monte Carlo simulations of the hadronic interactions connecting neutrino and \gr\ fluxes via the decay of charge and neutral pions created in $pp$ interactions within or close to a generic hadronic accelerator. This conversion relies on several assumptions on the source environment like sufficiently low radiation and matter densities to prevent \gr\ absorption, pion-matter and $p\gamma$ interactions. Details of these assumptions and considerations are given in~\cite{Kappes_NuGammaFlux}.

Unlike neutrinos, high energy \gr\ photons are absorbed by pair production on the extra-galactic background light (EBL). This process can be described by $\Phi_\mathrm{obs}=\Phi_\mathrm{source} \times e^{-\tau}$, where the optical depth $\tau$ is a function of the energy $E_\gamma$ and the redshift of the source $z_\mathrm{s}$. For sufficiently distant sources, i.e. sufficiently large optical depths, the expected gamma-ray flux $\Phi_\gamma$ will get absorbed and could therefore become compatible with the upper limits $\Phi_\mathrm{UL}$ derived from the H.E.S.S. measurements. Using the EBL model given in~\cite{EBL_Franceschini}, we derive these minimal distances to $z=0.007$ ($0.012$) for IC-5 and IC-18 respectively. It should be noted that this calculation very conservatively assumes that the potential sources are only emitting in the $100~\mathrm{TeV}$ range around the energies measured by IceCube, i.e. no lower energy radiation is present. Extrapolation of the IceCube flux to lower energies would allow to put severe constrains on the source distance, but would induce a significant dependence on the assumed spectral shapes.

\subsection{Real-time neutrino alerts}
Despite several efforts, including the ones presented here, the origin and astrophysical counterparts of the high-energy neutrinos detected by IceCube remain unknown. This hints to the possibility that the underlying sources are either very faint but numerous or of transient nature and emit neutrinos only for a limited amount of time. Should the latter be the case, the time domain of multi-messenger and multi-wavelengths searches becomes increasingly important. Only by triggering deep follow-up observations of significant neutrino events rapidly after their occurrence one can be sure to obtain complete contemporaneous multi-messenger and multi-wavelength coverage, which might be necessary for the unequivocal proof of a common origin of potentially observed transient events. 

The \hess\ multi-messenger program is actively pursuing this direction. Over the last years the second phase of the H.E.S.S. experiment has been commissioned successfully. It now includes the H.E.S.S.-II 28m telescope,  the largest Cherenkov telescope in the world and providing the lowest energy threshold of ground based gamma-ray detectors worldwide. Another major goal for this new phase of \hess\ is to reduce the response time of the system in order to increase the capabilities for the detection of transient phenomena. As discussed in~\cite{Hofverberg}, this goal has been achieved on the level of the instrument conception and construction. Significant further improvements to the DAQ and alert system have been introduced recently~\cite{Parsons_GRBs_ICRC2015}. The search for transient multi-messenger signals will greatly benefit from these developments. For example, close collaborations between \hess\ and the two major neutrino telescopes have been established and will enable rapid follow-up observations of high-energy neutrino candidates to start in the very near future.

\section{Gravitational waves}

\begin{figure*}[!t]
\vspace{-3mm}
\centering
\includegraphics[width=0.54\textwidth]{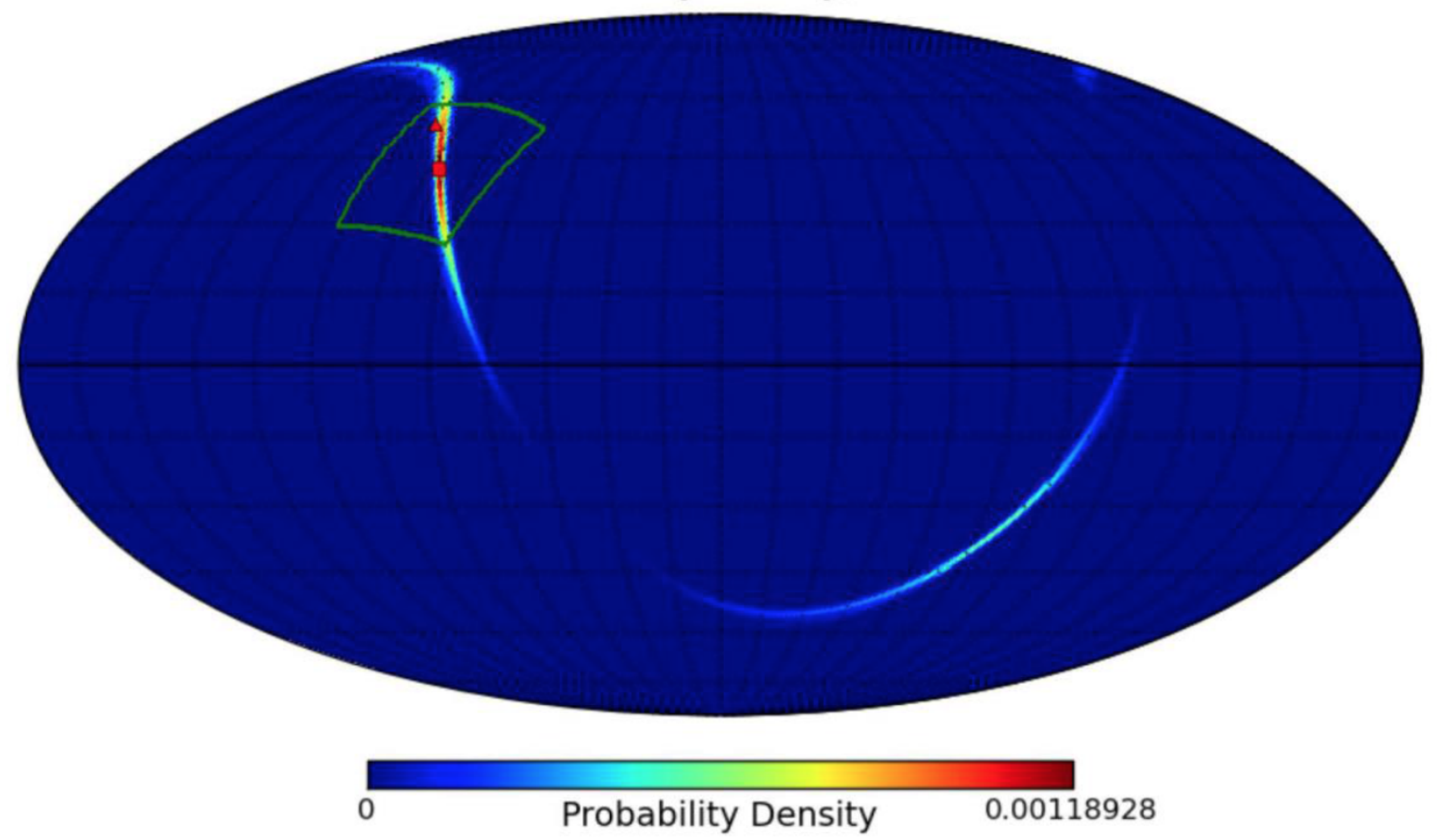}
\caption{Localization probability of the origin of a simulated gravitational wave detected jointly by the advanced Virgo and Ligo interferometers (from I. Bartos~\cite{BartosAMON14}).} \label{fig:GW}
\end{figure*}

In addition to the above-mentioned recent discoveries of extra-terrestrial, high-energy neutrinos, a completely new messenger might soon become available: gravitational waves. The two major, kilometer-scale interferometers searching for gravitational waves, Ligo and Virgo, are about to finish extensive upgrade programs and will start taking physics data in the next years (Ligo in 2015 and Virgo in 2016/2017). Predicted by the theory of General Relativity, direct detection of gravitational waves and the localization of their origin will permit important studies of both the underlying emission mechanisms as well as tests of fundamental principles of physics. The main signals for the advanced detectors are expected to come from the coalescence of binary neutron star systems. As these systems are also thought to be the progenitors for short gamma ray bursts, a connection between gravitational waves and other messengers like neutrinos, cosmic rays and gamma rays might be possible. 

Unfortunately the gravitational wave detectors will not be able to precisely localize the origin of the detected emission. Fig.~\ref{fig:GW} shows a typical example for the localization probability of the origin of a gravitational wave detected jointly by Virgo and Ligo. An extensive follow-up program involving a large range of instruments is therefore being set up. Thanks to their high sensitivity and large FoV, high-energy gamma ray observatories will play an important role in this endeavor (e.g.~\cite{BartosCTA}). The transient nature of the expected events (e.g. afterglows of short GRBs fade away within minutes) makes rapid follow-up responses a crucial requirement. Thanks to its rapid response time (combined with its high sensitivity and large FoV), the \hess\ \gr\ observatory is well suited for these observations and preparations for the first physics runs of the advanced GW detectors are in progress.

\section{Multi-frequency transients}
\hess\ continues to pursue its long-standing follow-up program on triggers from observatories covering the full electromagnetic spectrum from radio to high-energy \grs. For example, a dedicated program is performing follow-up observations on Gamma-Ray Bursts~\cite{Parsons_GRBs_ICRC2015}.\\
This program has been extended recently to include radio transients and especially Fast Radio Bursts (FRBs). Several of these extremely powerful and rapid bursts have been discovered in archival radio data over the last past years~\cite{FRBs}. They are considered to be emitted by extra-galactic objects but the underlying process remains elusive. Due to their brevity (lasting only a few milliseconds), online detections and rapid alerts to follow-up facilities play again a crucial role in the quest for their origin. Several radio observatories are currently putting in place such data analysis pipelines and first follow-up campaigns are being performed. \hess\ is currently collaborating with the SUPERB program at the Parkes radio observatory, able to emit FRB alerts with a delay O(20s)~\cite{Petroff}. 

\section{Acknowledgments}\vspace{-2mm}
\setstretch{0.8}
{\noindent{\scriptsize The support of the Namibian authorities and of the University of Namibia in facilitating the construction and operation of H.E.S.S. is gratefully acknowledged, as is the support by the German Ministry for Education and Research (BMBF), the Max Planck Society, the German Research Foundation (DFG), the French Ministry for Research, the CNRS-IN2P3 and the Astroparticle Interdisciplinary Programme of the CNRS, the U.K. Science and Technology Facilities Council (STFC), the IPNP of the Charles University, the Czech Science Foundation, the Polish Ministry of Science and Higher Education, the South African Department of Science and Technology and National Research Foundation, and by the University of Namibia. We appreciate the excellent work of the technical support staff in Berlin, Durham, Hamburg, Heidelberg, Palaiseau, Paris, Saclay, and in Namibia in the construction and operation of the equipment.}}


\end{document}